\documentclass[twocolumn]{aastex62}

\usepackage{hyperref}
\usepackage[]{amsmath}
\usepackage{soul}

\newcommand{\lambdacr}{\lambda_{cr}}
\newcommand{\taurec}{\tau_{rec}}
\newcommand{\tauAl}{\tau_{A,\lambda}}

\newcommand{\be}{\begin{equation}}
\newcommand{\ee}{\end{equation}}
\newcommand{\bea}{\begin{eqnarray}}
\newcommand{\eea}{\end{eqnarray}}

\newcommand{\Eq}[1]{Eq.~(\ref{#1})}

\graphicspath{{./}{figures/}}

\received{\today}
\revised{--}
\accepted{--}
\submitjournal{ApJ}

\shorttitle{Nonlinear reconnection in magnetized turbulence}
\shortauthors{Loureiro and Boldyrev}

\begin{document}

\title{\Large{Nonlinear Reconnection in Magnetized Turbulence}}

\correspondingauthor{Nuno F. Loureiro}
\email{nflour@mit.edu}

\author[0000-0001-9755-6563]{Nuno F. Loureiro}
\affil{Plasma Science and Fusion Center, Massachusetts Institute of Technology, Cambridge MA 02139, USA}

\author{Stanislav Boldyrev}
\affiliation{Department of Physics, University of Wisconsin at Madison, Madison, WI 53706, USA}
\affiliation{Space Science Institute, Boulder, Colorado 80301, USA}

\begin{abstract}
Recent analytical works on strong magnetized plasma turbulence have hypothesized the existence of a range of scales where the tearing instability may govern the energy cascade. 
In this paper, we estimate the conditions under which such tearing may give rise to full nonlinear magnetic reconnection in the turbulent eddies, thereby enabling significant energy conversion and dissipation. 
When those conditions are met, a new turbulence regime is accessed where reconnection-driven energy dissipation becomes common, rather than the rare feature that it must be when they are not.

\end{abstract}

\keywords{magnetic fields --- 
magnetic reconnection --- plasmas --- turbulence}

\section{Introduction} 
\label{sec:intro}
Magnetized plasmas are abundant in the Universe. 
Examples include the Earth's magnetosphere, the solar wind and the solar corona, as well as many others, more distant and often more exotic, such as accretion disks around massive central objects, astrophysical jets, pulsar wind nebulae, etc.  
Many such environments, including all those just listed, are turbulent --- a natural consequence of large-scale (roughly system-size) energy injection, and relatively infrequent collisions between the particles constituting those plasmas. 

Beyond its interest as a fundamental physics problem, an understanding of turbulence in those and other environments is believed to be crucial to address longstanding, fundamental processes such as dynamo action, enhanced loss of angular momentum in accretion disks,  electron-to-ion energy partition, and particle energization.

The modern understanding of (strong) plasma turbulence in the fluid approximation, though still incomplete, rests on a few qualitative ideas for which there is compelling observational and numerical evidence~\citep[e.g.][]{biskamp_magnetohydrodynamic_2003,chen2016,davidson_introduction_2016,schekochihin_mhd_2018}. 
Amongst these are: (i) a Kolmogorov-like cascade of energy from large to small scales; (ii) the concept of critical balance~\citep{goldreich_toward_1995} --- essentially a causality argument relating turbulent dynamics parallel and perpendicular to the local mean magnetic field; and (iii) the notion of dynamic alignment of the turbulent fluctuations~\citep{boldyrev_spectrum_2006,chandran_intermittency_2015,mallet_refined_2015}, which determines constraints imposed on the turbulence by the active alignment between velocity and magnetic field fluctuations. 

One direct consequence of the combination of these three concepts is the prediction that turbulent eddies should be anisotropic in all directions with respect to the local mean magnetic field; in particular, they should resemble current sheets --- localized regions of intense electric current --- in the field-perpendicular plane, whose aspect ratio increases with perpendicular wavenumber.  
Current sheets are, indeed, almost ubiquitously observed in direct numerical simulations of forced, three-dimensional magnetohydrodynamic (MHD) turbulence~\citep[e.g.][]{maron_simulations_2001,biskamp_magnetohydrodynamic_2003, zhdankin_statistical_2013}.

The extension of these ideas to the kinetic range of plasma turbulence --- relevant in weakly collisional plasmas of which the solar wind is the prototypical example --- is, predictably, not straightforward.
However, again, there is abundant numerical evidence for the formation of current sheets in this range~\citep[e.g.][]{tenbarge_current_2013,wan_intermittent_2015,groselj_fully_2018}.
Why this should be so has not been established on general theoretical grounds, but~\citet{boldyrev_role_2019} have recently advanced a possible explanation applicable to plasmas where $\beta_i\sim 1\gg \beta_e$, such as found, for example, in the Earth's magnetosheath (i.e., for so-called inertial kinetic-Alfv\'en wave turbulence~\citep{chen_nature_2017,passot_electron-scale_2017,passot_gyrofluid_2018,roytershteyn_numerical_2019}).

Current sheets being traditionally associated with magnetic reconnection~\citep{biskamp_magnetic_2000,priest_magnetic_2000,zweibel_magnetic_2009,yamada_magnetic_2010}, it is unsurprising that this process has acquired significant prominence as a potential key mechanism in magnetized turbulence~\citep[e.g.][]{matthaeus_turbulent_1986, retino_insitu_2007,sundkvist_dissipation_2007, servidio_magnetic_2009,osman_magnetic_2014,wan_intermittent_2015,cerri_plasma_2017,shay_turbulent_2018}. 
Fundamentally, reconnection leads to the conversion and dissipation of magnetic energy; thus, one expects that if it is indeed active in turbulence it may qualitatively impact the dynamics and observational signatures. 

An important point that needs to be introduced in our discussion at this stage is that of the relationship --- and distinction --- between magnetic reconnection and the tearing mode ~\citep{furth_finite-resistivity_1963}. 
The latter is an instability which manifests itself through the reconnection of magnetic field lines [and the consequent opening of magnetic islands (or flux ropes)]. Strictly speaking, therefore, the tearing mode {\it can} be called magnetic reconnection; however, the term ``reconnection'' is most commonly used to refer to a strongly nonlinear plasma phenomenon associated with significant magnetic energy transfer and dissipation, as already mentioned. 
According to this classification, the deep nonlinear stage of evolution of the (strongly unstable, i.e., large instability parameter $\Delta'$) tearing mode~\citep{coppi_resistive_1976, waelbroeck_current_1989,jemella_impact_2003,loureiro_x_2005} is appropriately referred to as reconnection; but not its linear and early nonlinear stages. 
Let us now see why this distinction matters.

Recent works~\citep{loureiro_role_2017, mallet_disruption_2017,boldyrev_magnetohydrodynamic_2017,loureiro_collisionless_2017,mallet_disruption_2017-1,loureiro_turbulence_2018,boldyrev_calculations_2018, boldyrev_role_2019} have presented analytical arguments for the inevitability of the onset of the tearing mode (in either its resistive or collisionless forms, as appropriate) below a certain (so-called, critical) turbulence scale, $\lambdacr \ll L$, where $L$ is the energy injection scale, in a wide variety of plasma regimes. 
It is argued by these authors that the effect of the tearing mode is to redefine the energy cascade rate (to become the tearing mode growth rate, see Section~\ref{sec:prelim}), resulting in a different energy spectrum and eddy anisotropy at scales $\lambda \ll \lambdacr$. 
Magnetohydrodynamic simulations performed subsequently appear to lend support to these ideas~\citep{walker_influence_2018,dong_role_2018}, as does a detailed analysis of solar wind data~\citep{vech_magnetic_2018}.
Additional consistent numerical evidence has been reported by~\citet{arzamasskiy_hybrid-kinetic_2019,landi_spectral_2019} (specifically, the measurement of linear anisotropy of the turbulent fluctuations, $k_\parallel\sim k_\perp$, in the sub-ion range, as predicted for tearing-mediated inertial kinetic-Alfv\'en wave turbulence~\citep{boldyrev_role_2019}).

Tearing onset in turbulence thus appears to be in reasonably strong footing --- prompting the important question of whether it can (and, if so, under what conditions) lead to a fully nonlinear reconnecting stage. 
Essentially, the reason this question is non-trivial is that the reconnection rate differs from the tearing mode growth rate (and, therefore, from the eddy turnover rate in the tearing-mediated turbulence range). 
The goal of this paper is to address this issue\footnote{Hopefully, the reason for the somewhat tautological title of this paper is now clear. 
Reconnection is usually implicitly understood to be a nonlinear phenomenon.
The specific phrasing of the title aims to stress the distinction between the linear and early nonlinear evolution of the tearing mode, on one hand, and its late, strongly nonlinear evolution on the other --- the latter being what is meant here by the proper, or nonlinear, reconnection stage. This distinction is key, since the linear and early nonlinear stages of the tearing mode reconnect insignificant amounts of flux, and lead to negligible energy dissipation and conversion.}. 

\section{Preliminaries}
\label{sec:prelim}
The key idea underlying the suggestion that the tearing mode is activated at turbulence scales $\lambda \ll \lambdacr$ derives from the observation that, at such scales, the tearing mode growth rate, $\gamma_t(\lambda)$, exceeds the eddy turn-over rate, $\tau_{nl}^{-1}(\lambda)$, that would otherwise pertain to those scales, i.e.,
\begin{equation}
\label{eq:tear_onset}
    \gamma_t\tau_{nl}\gg 1,
\end{equation} 
with $\lambdacr$ resulting from solving this condition in the case of approximate equality~\citep{loureiro_role_2017,mallet_disruption_2017}. 
It is demonstrated in these references that the specific mode (wavenumber) that solves~\Eq{eq:tear_onset}, amongst all possible tearing-unstable modes, is the fastest-growing tearing mode (often dubbed the `Coppi' mode~\citep{coppi_resistive_1976}).

The onset of the tearing mode, {\it per se}, is not sufficient to interfere with the turbulent cascade. Its ability to be dynamically significant naturally hinges on whether it can attain a nonlinear amplitude. 
In this regard, the tearing mode is a somewhat peculiar instability in that it becomes nonlinear at very small amplitudes: i.e., as soon as the width of the magnetic island that it creates exceeds the thickness of the inner boundary layer (which is, forcefully, asymptotically smaller than the characteristic length scale of variation of the background magnetic profile; i.e., in this case, than the size of the eddy, $\lambda$).\footnote{To be specific: purely from geometric considerations and the definition of separatrix, one has that the full width of a magnetic island is given by $W=4\sqrt{-\tilde\psi/\psi_{eq}''}$, where $\tilde \psi$ is the perturbed (reconnected) flux, and $\psi_{eq}''\approx B_{eq}/a$ is the equilibrium current at the rational layer. The tearing mode becomes nonlinear when $W\approx \delta_{in}$, where $\delta_{in}$ is the width of the inner boundary layer that arises in the tearing mode calculation~\citep{furth_finite-resistivity_1963,coppi_resistive_1976}; it scales with resistivity in the MHD regime, or with electron inertia in kinetic calculations and is, by definition, asymptotically smaller than $a$, the width of the background equilibrium. One thus finds that, in the early non-linear regime, $\tilde \psi/\psi_{eq}\approx 1/16\, (\delta_{in}/a)^2\ll 1$; note however that, numerical pre-factor aside, this condition implies that, for the Coppi mode, the perturbed and background currents are comparable at this stage.}
As the tearing mode begins its nonlinear evolution, it continues to grow exponentially at the same rate as in the linear stage\footnote{This is not generally true for all wavenumbers unstable to tearing, but it is true for the most unstable (Coppi) mode.}~\citep{wang_nonlinear_1993, porcelli_recent_2002, loureiro_x_2005}.
These notions imply that ~\Eq{eq:tear_onset} correctly represents the condition for the nonlinear tearing mode to affect the turbulent cascade \cite[][]{loureiro_role_2017,boldyrev_magnetohydrodynamic_2017,mallet_disruption_2017}. Furthermore, the tearing mode onset implies that $\gamma_t$ becomes the eddy turnover rate at those scales, with a consequent change in the turbulence spectrum and other properties.

However, and as we now explain, it is less clear --- but, we will argue, critical --- whether the (early) nonlinear stage of the tearing mode evolution has the chance to evolve towards the deep nonlinear (i.e., properly reconnecting) stage, whereupon a significant amount of the magnetic flux in the eddy is reconnected, and considerable magnetic energy dissipation and conversion occurs. 

In the absence of background turbulence, the (strongly unstable, large $\Delta'$) tearing mode is known to transition to a fully nonlinear reconnecting state once its amplitude becomes sufficiently large~\citep{waelbroeck_current_1989,jemella_impact_2003,loureiro_x_2005,loureiro_fast_2013}. 
At this moment, the tearing rate will, in most cases, change to a different value, usually referred to as the (normalized) reconnection rate, $\mathcal{R}$.
The current understanding of reconnection suggests the following.
In resistive MHD, there are two possibilities for $\mathcal{R}$, depending on the value of the Lundquist number, $S=L_{CS} v_A/\eta$, where $L_{CS}$ is the current sheet length, $v_A$ the Alfv\'en speed based on the reconnecting component of the magnetic field, and $\eta$ is the magnetic diffusivity. 
If $S\lesssim S_{cr}\approx 10^4$ we have $\mathcal{R}=S^{-1/2}$ (i.e., the Sweet-Parker rate~\citep{parker_sweets_1957,sweet_neutral_1958}). 
This is the only case where, in fact, the reconnection rate is the same as the tearing rate (of the most unstable tearing mode). 
However, this result is of limited applicability as, generally, $S\gg S_{cr}$; in such cases, one instead has 
$\mathcal{R}=S_{cr}^{-1/2}\approx 0.01$~\citep{loureiro_instability_2007,samtaney_formation_2009,bhattacharjee_fast_2009,huang_scaling_2010,uzdensky_fast_2010, loureiro_magnetic_2012,loureiro_magnetic_2016}.
For collisionless reconnection, though absent a theoretical explanation (see, however,~\citet{liu_why_2017}), it is generally accepted that $\mathcal{R}\approx 0.1$~\citep{birn_geospace_2001,comisso_value_2016, cassak_review_2017}.

For use in what follows, let us define the reconnection time as
\begin{equation}
\label{eq:tau_rec}
    \taurec=\mathcal{R}^{-1}\tauAl,
\end{equation}
where $\tauAl=\lambda/v_{A,\lambda}$. 
The physical meaning of $\taurec$ is that it is the time that it takes to reconnect the magnetic flux contained in an eddy of size $\lambda$ and reconnecting field $B_\xi(\lambda)$.
The question which we wish to address is whether this reconnection rate is larger than the tearing rate, that is, whether an eddy distorted by the tearing instability may end up reconnecting significant magnetic flux,  thus leading to significant energy conversion and dissipation.
We propose that this will only happen if
\begin{equation}
    \label{eq:rec_onset}
    \gamma_t\taurec\ll 1.
\end{equation}
In other words, a typical eddy at scales $\lambda\gg \lambdacr$ exists for a time of order $\gamma_t^{-1}$.
It, therefore, has a finite probability of reaching the deep nonlinear stage, whereupon it may transition to the reconnection regime. If condition~(\ref{eq:rec_onset}) is met, then the reconnection time is much shorter than the eddy turnover time, and it is thus expected that full reconnection will occur. 
Otherwise, reconnection is slower, and the eddy will cease to exist without significant reconnection having taken place.
We now proceed to compute this condition, and discuss its implications, in three different cases: the pure MHD case, Section~\ref{sec:mhd}; and the cases when tearing, and reconnection, are enabled by kinetic physics (electron inertia) instead of resistivity, and the eddies in which they happen are above (Section~\ref{sec:fluid}) or below (Section~\ref{sec:kin}) the ion kinetic scales.
\section{The magnetohydrodynamic case}
\label{sec:mhd}
The onset of tearing in MHD turbulence has been addressed by ~\citet{loureiro_role_2017,mallet_disruption_2017,boldyrev_magnetohydrodynamic_2017}.
These authors find
\begin{equation}
    \lambdacr/L\sim S_L^{-4/7},
\end{equation}
where $S_L=L V_{A,0}/\eta$ is the outer scale Lundquist number, and $V_{A,0}$ is the Alfv\'en velocity based on the background (mean) field $B_0$.
Below this scale, the eddy turnover time becomes the growth rate of the fastest growing tearing mode:
\begin{equation}
    \gamma_t\sim \tauAl^{-1}(\lambda v_{A,\lambda}/\eta)^{-1/2}.
\end{equation}
where~\citep{boldyrev_magnetohydrodynamic_2017}
\begin{equation}
\label{eq:valambda}
    v_{A,\lambda}\sim\varepsilon^{2/5}\eta^{-1/5}\lambda^{3/5},
\end{equation}
with $\varepsilon=V_{A,0}^3/L$ the injected power.

Therefore, evaluation of \Eq{eq:rec_onset} yields the requirement
\begin{equation}
\label{eq:mhd_requirement}
    \lambda/L \gg \mathcal{R}^{-5/4}S_L^{-3/4}.
\end{equation}

It is necessary for the validity of this result that $\lambdacr\gg \lambda \gg \lambda_{diss}$, where $\lambda_{diss}\sim S_L^{-3/4}L$~\citep{boldyrev_magnetohydrodynamic_2017} is the dissipation scale. 
Since $\mathcal{R}<1$, the second inequality is automatically satisfied. 
As to the first, we find that it implies
\begin{equation}
\label{eq:cond_Lund}
    S_L\gg\mathcal{R}^{-7}.
\end{equation}
As mentioned before, as long as $S_\xi=\xi v_{A,\lambda}/\eta\gtrsim S_{cr}\sim 10^4$,\footnote{To check whether this is true, we use~\Eq{eq:valambda} and the scaling $\xi\sim L (\lambdacr/L)^{1/4}(\lambda/\lambdacr)^{9/5}$~\citep{boldyrev_magnetohydrodynamic_2017}, both of which expressions are valid in the tearing-mediated turbulence range. Then, we find that $S_\xi\gg S_{cr}\implies \lambda/L\gg S_{cr}^{5/12}S_L^{-73/84}$. This is smaller than $\lambdacr/L$ if $S_L\gg S_{cr}^{7/5}\sim 10^{28/5}$. This condition is superseded by~\Eq{eq:cond_Lund}.} the reconnection rate is $\mathcal{R}\sim S_{cr}^{-1/2}\sim 0.01$. 
We thus arrive at the conclusion that significant reconnection is only possible if $S_L\gg 10^{14}$, a considerable demand even by the standards of astrophysical and space plasmas --- and certainly  one that direct numerical simulations cannot be imagined to meet anytime in the foreseeable future\footnote{It is straightforward to repeat this calculation for the case when the profile of the reconnecting magnetic field in the eddy is better represented by $\sin(x)$, instead of the default Harris ($\tanh(x)$) profile that we consider here. 
In that case, $\gamma_t\sim \tauAl^{-1} (\lambda v_{A,\lambda}/\eta)^{-3/7}$, with $v_{A,\lambda}\sim \varepsilon^{7/18}\eta^{-1/6}\lambda^{5/9}$~\citep{boldyrev_magnetohydrodynamic_2017}.~\Eq{eq:mhd_requirement} is then replaced by $\lambda/L\gg \mathcal{R}^{-3/2} S_L^{-3/4}$.
The estimation of the dissipation scale is unchanged from that pertaining to the Harris sheet; but, in this case, $\lambdacr/L\sim S_L^{-6/11}$. 
Therefore, \Eq{eq:cond_Lund} is instead $S_L\gg \mathcal{R}^{-22/3}$.}.

\section{The collisionless case}
\label{sec:collisionless}
Let us now examine the same question in a plasma where collisions are sufficiently rare that the breaking of frozen flux condition required to enable the tearing mode and the subsequent nonlinear reconnection stage is due to electron inertia (active at the electron skin-depth scale, $d_e=c/\omega_{pe}$), rather than resistivity as considered in the previous section; i.e., in this case $\mathcal R \approx 0.1$.

As documented in ~\citet{loureiro_collisionless_2017}, there are two cases that need considering: the first, somewhat simpler to address, is when the critical scale at which the tearing mode onsets is in the MHD range (i.e., $\lambdacr$ is larger than the ion kinetic scales) --- eventhough, to repeat, the tearing and reconnection themselves require kinetic effects. 
This is treated in Section~\ref{sec:fluid}.
The second case is when the onset of tearing only occurs for scales smaller than the ion kinetic scales. 
This is discussed in Section~\ref{sec:kin}.
\subsection{Reconnection at fluid scales}
\label{sec:fluid}
Several cases are possible, depending on plasma parameters~\citep{loureiro_collisionless_2017}. 
There is no need here to be exhaustive: for any particular case, the calculation proceeds in a qualitatively similar way. 
Therefore, let us consider, as an example, a low beta plasma (see Sections 2 \& 3 of~\citet{loureiro_collisionless_2017}) --- we choose to analyze this particular case because of its potential relevance to solar wind observations~\citep{vech_magnetic_2018}, and perhaps also to the solar corona.
In this case, in the tearing-mediated range, the eddy turnover rate becomes the growth rate of the fastest growing tearing mode as given by $\gamma_t\sim v_{A,\lambda} d_e\rho_s/\lambda^3$, where $\rho_s$ is the ion sound Larmor radius.
Evaluation of \Eq{eq:rec_onset} then yields:
\begin{equation}
\label{eq:fluid_colless}
    \lambda\gg\mathcal{R}^{-1/2}(d_e\rho_s)^{1/2}.
\end{equation}
This expression only applies in the range of scales $\lambdacr\gg \lambda \gg \rho_s$ where, for this case, $\lambdacr/L\sim (d_e/L)^{4/9}(\rho_s/L)^{4/9}$~\citep{loureiro_collisionless_2017,mallet_disruption_2017-1}; this translates into
\begin{equation}
    \label{eq:derhos}
    \mathcal{R}\ll \frac{d_e}{\rho_s}\ll\mathcal{R}^9\left(\frac{L}{\rho_s}\right)^2.
\end{equation}
The left inequality is probably not satisfied in the (pristine) solar wind at $\sim 1$ AU (it requires $\beta_e\ll 2 (m_e/m_i)\mathcal{R}^{-2}\approx 0.1$, which may be  too low).
In that case, one concludes that reconnection in current sheets should not be a main energy dissipation mechanism in that turbulent environment.
The opposite situation, however, should pertain to the solar corona: 
using standard parameters there is no difficulty in concluding that both inequalities in Eq.~(\ref{eq:derhos}) should hold comfortably.\footnote{Note that this is indeed the regime that we would expect to describe turbulence in the solar corona at these scales, rather than the MHD case of Section~\ref{sec:mhd}: for typical coronal conditions, i.e., $S_L\approx 10^{14}$, $L\approx 10^4\,$km, Eq. (15) of~\citet{loureiro_role_2017}, describing the MHD case, yields an estimate of the width of the inner boundary layer of the tearing mode occurring on an eddy of width $\lambdacr$ of approximately $1\,$cm, smaller than the electron skin depth and showing, therefore that the tearing mode at such scales is collisionless, as we consider in this Section.} 
From the point of view of numerical simulations, this result, like \Eq{eq:cond_Lund}, unfortunately places close to impossible demands\footnote{For the $\sin(x)$ profile, \Eq{eq:fluid_colless} becomes instead $\lambda\gg\mathcal{R}^{-2/3}d_e^{4/9}\rho_s^{5/9}$; whereas \Eq{eq:derhos} is replaced by $\mathcal{R}^{3/2}\ll d_e/\rho_s\ll \mathcal{R}^{63/6} (L/\rho_s)^{9/4}$. The conclusions drawn above pertain equally for these estimates.}.

\subsection{Reconnection at kinetic scales}
\label{sec:kin}
Finally, we analyze the case when the tearing onset only occurs at scales below the ion kinetic scales. Let us consider here the analysis recently proposed by~\citet{boldyrev_role_2019} of sub-ion range turbulence in plasmas such that $\beta_i\sim 1\gg\beta_e$. 
The relevant eddy turnover rate is
\begin{equation}
\label{eq:kin_gamma}
    \gamma_t\sim \frac{V_{Ae,\lambda}}{\lambda}\left(\frac{d_e}{\lambda}\right)^2.
\end{equation}
Therefore we find:
\begin{equation}
\label{eq:kinetic}
    \gamma_t\taurec \ll 1 \implies\frac{\lambda}{d_e} \gg \mathcal{R}^{-1/2}\sim 3.
\end{equation}
Repeating this derivation for the $\sin(x)$ magnetic field profile instead yields $\lambda/d_e\gg \mathcal{R}^{-2/3}\sim 5$.

Unlike the two cases considered previously, an estimation of $\lambdacr$ is not available for this situation (a reflection of the fact that a detailed understanding of sub-ion range turbulence is still lacking). 
The only known constraint that applies to $\lambdacr$ is that it be smaller than $\min(d_i, \rho_i,\rho_s)$, which simply follows from the range of validity of the equations that are used by~\citet{boldyrev_role_2019} to compute~\Eq{eq:kin_gamma}. 
At small scales, it is required that $\lambda\gg d_{e}$ which is (marginally) satisfied by~\Eq{eq:kinetic}.

It is interesting to analyze this result in light of recent MMS observations of so-called electron-only reconnection in the Earth's magnetosheath~\citep{phan_electron_2018,stawarz_properties_2019}. 
In~\citet{boldyrev_role_2019} we have estimated that the decoupling of the ions in these events requires $\lambda/d_e\ll\sqrt{d_i/d_e}$ or  $\lambda/d_e\ll{(d_i/d_e)}^{2/3}$ depending on whether one assumes a $\tanh(x/\lambda)$ or $\sin(x/\lambda)$ magnetic field profile for the reconnecting field $B_\xi(x)$ in the eddy.  
These estimates range from $\sim 6$ to $\sim 12$ suggesting, therefore, a rather narrow range of scales where nonlinear `electron-only' reconnection in the eddies may be possible. 
Remarkably, \citet{phan_electron_2018} report a current sheet thickness of $\sim 4\, d_e$, strikingly consistent with these numbers and with \Eq{eq:kinetic}.
This is certainly very encouraging, but one must also bear in mind that all our analytical results are only order of magnitude estimates which ignore order unity numerical prefactors.

Another observationally-based result which we interpret to be consistent with our analysis is the recent claim by~\citet{chen_evidence_2019} that energy dissipation at kinetic scales in the magnetosheath is dominated by linear electron Landau damping (the energy dissipation rate via that channel being comparable to the energy cascade rate).
Indeed, \Eq{eq:kinetic} demonstrates that full reconnection in sub-ion scale eddies is permitted for typical magnetosheath parameters; and previous investigations of heating in (strong guide-field) collisionless reconnection~\citep{loureiro_fast_2013,numata_ion_2015} show that when $\beta_e\ll 1$ linear electron Landau damping is by far the dominant energy dissipation channel.
\section{Conclusion}
\label{sec:conclusion}
This paper builds on previous recent work on the onset of the tearing instability in strong magnetic plasma turbulence, establishing the conditions under which this instability may develop into a deep nonlinear reconnecting state. 
The ability to do so is intimately tied to whether or not significant energy dissipation and conversion is to be expected at such turbulent scales. 
We think this has profound implications for turbulent systems. 
For example, in weakly collisional plasmas, reconnection is a well known efficient particle acceleration mechanism~\citep[e.g.][]{guo_formation_2014,Sironi_relativistic_2014,dahlin_electron_2015, werner_non-thermal_2017}, and heats different species at different rates~\cite[e.g.][]{numata_ion_2015,shay_turbulent_2018}.
Therefore, if reconnecting eddies are a common occurrence --- the conditions for which are worked out in this paper --- then one might expect turbulence to be efficient at generating non-thermal populations and different electron-to-ion temperature ratios, which are indeed observed or expected in different space and astrophysical plasmas~\citep[see, e.g.,][and references therein]{schekochihin_constraints_2019}.
Moreover, the very observability of reconnecting turbulence depends, obviously, on whether truly reconnecting eddies are the norm or an exception.

Yet another consequential implication of the analysis carried out in this paper stems from the fact that neither~\Eq{eq:cond_Lund} nor \Eq{eq:derhos} have ever been met in computer simulations conducted to date, nor is that likely to happen in the near future.
It thus follows that {\it all} observations of reconnecting current sheets in (three-dimensional) numerical simulations of strong turbulence in the plasma regimes to which those equations pertain are bound to be relatively rare or transient events, with no significant impact on the nature of energy dissipation. 
One immediate consequence is, therefore, that the energy dissipation (heating or particle energization) rates obtained in simulations of magnetic turbulence may be severely underestimated with respect to the environments that such simulations aim to study. 
One way to remedy this situation might be to hardwire, in numerical simulations, energy dissipation prescriptions based on the energetics of reconnection (specific to the particular plasma parameters under study) at scales where the turbulent cascade is tearing dominated.

NFL was partially
funded by NSF CAREER award no. 1654168 and by
the NSF-DOE Partnership in Basic Plasma Science and
Engineering, award no. de-sc0016215. 
SB was partly supported by the NSF under grant no. NSF PHY-1707272,
by NASA under grant no. NASA 80NSSC18K0646, and
by DOE grant no. DE-SC0018266.



\end{document}